\documentclass{aa}
\usepackage{graphicx}

\begin{document}

\headnote{Research Note}
\title{Is the Giant Elliptical Galaxy NGC~5018 a Post-Merger Remnant?}

\author{L.~M.~Buson\inst{1}, F.~Bertola\inst{2}, A.~Bressan\inst{1},
D.~Burstein\inst{3} and M.~Cappellari\inst{4}}

\offprints{Lucio M. Buson, \email{buson@pd.astro.it}}

\institute{
INAF Osservatorio Astronomico di Padova, vicolo
dell'Osservatorio~5, I-35122 Padova, Italy\and
Dipartimento di Astronomia, Universit\`a di Padova, vicolo
dell'Osservatorio~2, I-35122 Padova, Italy\and
Department of Physics \& Astronomy,
Arizona State University, Tempe, AZ, 85287-1504, USA\and
Leiden Observatory, Postbus 9513, 2300 RA Leiden, The Netherlands}

\date{Received ... / Accepted ... }

\titlerunning{NGC 5018}
\authorrunning{Buson et al.}

\abstract{NGC~5018, one of the weakest UV emitters among giant
ellipticals (gE) observed with IUE, appears to consist of an optical
stellar population very similar to that of the compact, dwarf
elliptical M32, which is several magnitudes fainter in luminosity than
NGC~5018 and whose stellar population is know to be $\sim$ 3 Gyr
old. Here we show that the mid-UV spectra of these two galaxies are
also very similar down to an angular scale hundreds times smaller than
the IUE large aperture (as probed by HST/FOS UV spectra obtained
through 0.86$''$ apertures). This implies a reasonably close match of
the populations dominating their mid-UV light (namely, their
main-sequence turnoff stars). These data indicate that NGC~5018 has,
in its inner regions, a rather uniform dominance of a $\sim 3$ Gyr-old
stellar population, probably a bit different in metallicity from M32.
Combined with the various structures that indicate that NGC~5018 is
the result of a recent major merger, it appears that almost all of
stars we see in its center regions were formed about 3 Gyr ago,
in that merger event. NGC~5018 is likely the older brother of NGC~7252,
the canonical gE-in-formation merger. As such, NGC~5018 is perhaps
the best galaxy which can tell us how a merger works, after the
fireworks subside, to form a gE galaxy today. For this reason alone,
the stellar populations in NGC~5018 at all radii are worth studying
in detail.
}

\maketitle

\section{Introduction}

NGC~5018 is the dominant giant elliptical of a small group
(Gourgolhon et al. 1992), whose distance is kinematically estimated
between 40.8~Mpc (Faber et al. 1989) and 39.9~Mpc (Hyperleda
Database; Paturel et al. 2003). An independent, almost identical
estimate ($\sim$39.8~Mpc) comes also from the observed light curve
of its recent SN~Ia 2002dj (Hutchings \& Li 2002), having adopted a
total reddening E(B$-$V)$\sim$0.2 along its line of sight (G.~Pignata,
private communication).

NGC~5018 is quite peculiar in two main respects. First,
Schweizer et al (1990) classified NGC~5018 as one its prime
candidates for a recent major merger, assigning it a $\Sigma$
parameter of 5.15. Second, although it is morphologically classified as a
gE, its nuclear optical spectrum distinguishes itself by
having the weakest measurement of the absorption line index Mg$_2$
for its velocity dispersion among the over 400 gEs surveyed by the
7Samurai ($\rm Mg_2 = 0.218 \pm 0.007$; Davies et al. 1987) .
Its unusual properties brought it to our attention in the early
1990's, and the IUE spectrum we obtained (Bertola et al. 1993) lacks
the prominent UV-upturn shortward of $\lambda$~2000\AA, that is
typical of old, metal-rich spheroids (Burstein et al. 1988).

Given these data, it is not surprising that some observers
have suggested the existence of a fine-tuned conspiracy, where
young stars and heavy dust absorption conspire to dilute the
underlying Mg$_2$ line strength index, also turning a flat,
young-star-dominated UV energy distribution into the observed NGC~5018
UV-weak spectrum (Carollo \& Danziger 1994). However, this scenario
is not consistent with the finding that the average internal reddening
is as low as E(B$-$V)$\sim$0.02, within the 10$''$$\times$20$''$ IUE
aperture (Buson et al. 2001). In the present Research Note
we re-examine the stellar population of NGC~5018 by comparing its
UV spectra (obtained with both HST/FOS and IUE) with similar spectra
of M32. In so doing, we believe we can now identify NGC~5018 as
the best candidate we have for the result of a major merger forming
a gE galaxy in today's universe.

\section{The UV Evidence}

Bertola et al. (1993) found an unexpectedly good match between
the far-UV spectra of NGC~5018 (M$_B$=-21.52; $\sigma$=212 km
s$^{-1}$; Prugniel \& Simien 1996) and of the cE dwarf galaxy M32,
four magnitudes less luminous (M$_B$=-17.76; $\sigma$=74 km
s$^{-1}$). Such a similarity includes the lack of far-UV
flux, comparable mid-UV spectral shape, as well as similar
Mg$_2$ line-strength index values in the optical (0.196$\pm$0.007; for
M32; Trager et al. 1998, compared to 0.218 for NGC~5018).  The HST/FOS
spectra we discuss below show that mid-UV homogeneity of the two
galaxy populations holds even at subarcsec angular scales. Any
viable interpretation one wants to give to the origin of the
stellar population of NGC~5018 must account for the rather uniform
stellar population that is seen throughout its central regions.

\begin{figure}
\resizebox{\hsize}{!}{\includegraphics{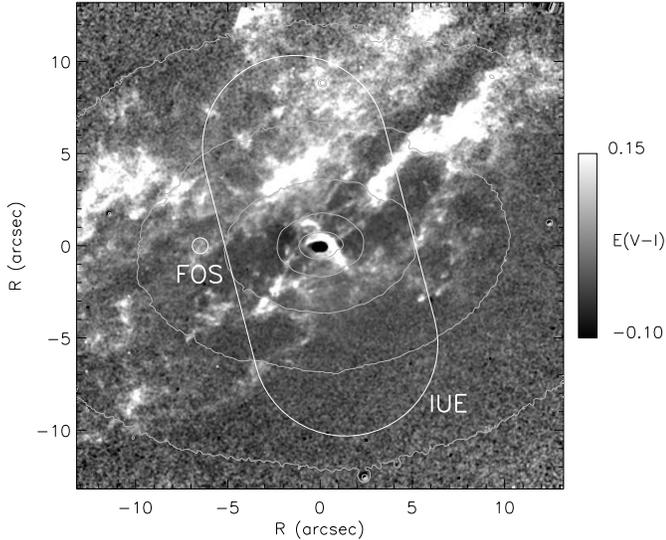}}
\caption{Color excess map $E(V-I)$ of the center of NGC~5018, as
obtained from HST/WFPC2 observations, after subtracting the underlying
stellar color gradient of Fig.~\ref{fig:fos_coo}. Superimposed on this
field are the locations of the IUE aperture (approximately
10$''\times$20$''$ in size), as well as of the 0.86$''$ diameter
circular aperture used for the HST/FOS off-nucleus observation. The
gray contours levels represent the $I$-band galaxy surface brightness,
in steps of 1 mag arcsec$^{-2}$. North is up and east is to the left.}
\label{fig:fos_loc}
\end{figure}

\begin{figure}
\resizebox{\hsize}{!}{\includegraphics{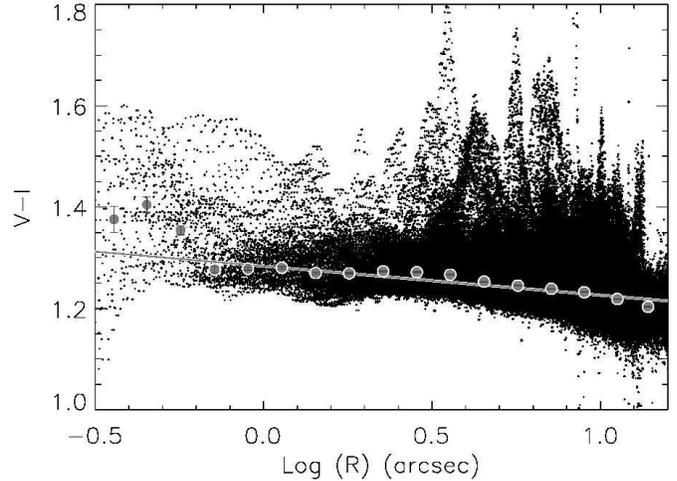}}
\caption{Determination of the dust-corrected central $V-I$ color
gradient from the HST/WFPC2 images, using the method of Cappellari et
al. (2002).  The black dots represent the color of the individual
pixels in the PC1 CCD, as a function of their elliptical radius. The
straight line is a robust fit which minimizes the sum of the absolute
deviations of the points from the line, being thus almost insensitive
to dust effects. The best fit, calibrated to the Johnson system
following Dolphin (2000) gives $(V-I)$ = (1.283$\pm$0.004) $-$
(0.057$\pm$0.006)*log(r/1$''$), its zeropoint being not corrected for
galactic reddening. Filled circles represent the median color
computed in logarithmically spaced radial bins (formal errors are
smaller than the symbol size).}
\label{fig:fos_coo}
\end{figure}

\begin{figure*}
\centering
\includegraphics[width=15cm]{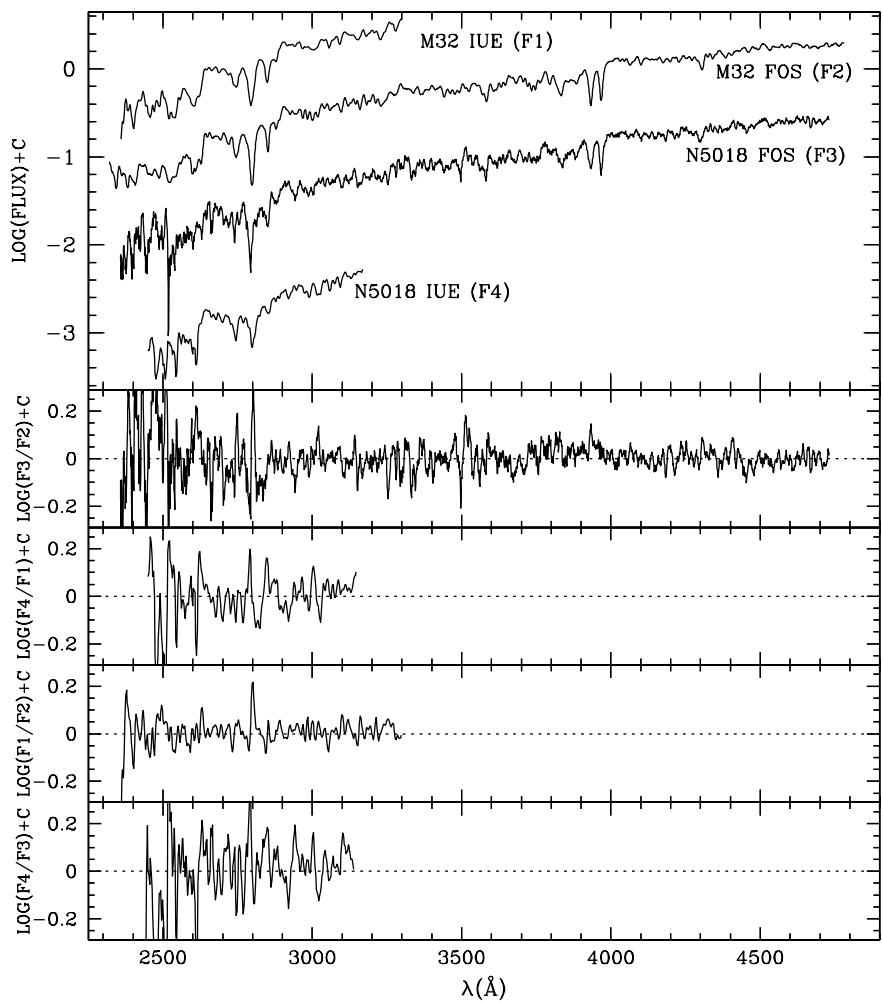}
\caption{Top panel: IUE spectra and FOS G270H$+$G400H spectra for the
center region of M32 and the bulge of NGC~5018. Lower panels: the
divisions of each spectrum by each other, namely M32/FOS by
NGC5018/FOS, M32/IUE by NGC5018/IUE, M32 FOS/M32 IUE and NGC5018
FOS/NGC5018 IUE. All plots are given in either log F$_\lambda$$+$C or
log [F$_\lambda$ (X)/F$_\lambda$(Y)]$+$C {\em vs.} $\lambda$.}
\label{fig:fos_alo}
\end{figure*}

Leonardi \& Worthey (2000) proposed that NGC~5018 is an old,
metal-rich elliptical which has undergone a major accretion event at
some recent stage of its evolution.  In this interpretation, NGC~5018
appears dominated by a stellar population similar to that of M32,
about 3 Gyr-old from current estimates (Trager et al. 2000; Terlevich \&
Forbes 2002), which masks the old stellar population in this galaxy.

Interest in the puzzling stellar population content of NGC~5018 pushed
one of us (DB) to include it in a HST-based spectroscopic survey carried
out through the (circular) Faint Object Spectrograph 0.86$''$ aperture
(Ponder et al. 1998). This dataset included four
exposures with the UV grating G270H
(4$\times$2400~s) and a single exposure (1020~s) with the UV/optical
grating G400H, giving continuous spectral coverage from 2300 to 4800~\AA.
Luckily, the HST archives also contain an analogous FOS
G270H$+$G400H set of spectra taken with the 0.86$''$ square FOS
aperture of the center of M32 (programme ID=6636; PI: M.Gregg). It
also consists of four G270H exposures (4$\times$1000~s) and an
additional, short-exposure (375~s) G400H spectrum. The net result is
that we fortuitously have rare tools to probe directly the stellar
content of these two galaxies over a wide range of angular scales.

In the case of NGC~5018 the FOS aperture was located off-nucleus, the
result of an initially incorrect telescope position. As such, this UV
spectrum corresponds to a region of the galaxy body outside the oval
aperture of the previous IUE spectrum (though within the galaxy's
effective radius r$_e$$\sim$15$''$; Alonso et al. 2003). Its final
position, including also the fine adjustment due to the ACQ/PEAK
pointing procedure, is $\sim$6$''$ east of the nucleus, as shown in
Fig.~\ref{fig:fos_loc}. The latter image represents the
$E(V-I)$ color excess map of the center of NGC~5018, obtained by
calibrating properly aligned, deep HST/WFPC2 archive observations of
F. Schweizer (four, 1100-s F555W plus four, 1300-s F814W exposures),
once the underlying stellar $(V-I)$ color gradient shown in
Fig.~\ref{fig:fos_coo} has been subtracted. The total eastward FOS
aperture shift has been independently cross-checked by comparing
the FOS spectrum level with the flux recorded at the same position
in a UV WFPC2/F336W frame taken from a previous HST programme of
ours (ID 5732).

In addition to these HST/FOS observations, IUE also observed both
NGC~5018 and M32. The spectra discussed here consist of single,
low-resolution LWP exposures obtained through the large, oval
aperture centered on their nuclei and oriented at P.A=14$^\circ$ and
P.A=153$^\circ$, respectively. The corresponding exposure times
are 335 and 180~min.

\section{Results}

The major outcome of these comparisons between the
two galaxies is that the match (Fig.~\ref{fig:fos_alo}) of
both the overall SED and the mid-UV triplet, {\em i.e.} of the
absorption lines formed by the Mg~II 2800~\AA, Mg~I 2852~\AA\ and the
Fe~I $+$~II $+$~Cr~I feature at 2750~\AA, well-known for showing rapid
changes in relative strengths from late B to late F stars, holds
over a wide range of angular sizes, from 10's of arc-seconds
down to a $<$~1$''$ field of view. As far as FOS spectra are concerned,
a more quantitative appreciation of the match of NGC~5018 and
M32 SEDs shown in Fig.~\ref{fig:fos_alo} can be obtained by the standard
mid-UV indices listed in Table~\ref{tab:fos_ind}, measured after
a proper velocity dispersion match of the original spectra. (Note:
the apparent inversion of the Ca II H\&K lines in the FOS spectrum
of NGC~5018 is most likely due to an unfortunate cosmic ray, as this
inversion does not correspond to the stellar population seen in
the other parts of this spectrum.)

\begin{table}
\caption{FOS Ultraviolet Spectral Indices}
\centering
\begin{tabular}{lcccc}
\hline\hline
Feature & M32 & Error & N5018 & Error\\
\hline
2600-3000  & 1.459 & 0.003 & 1.340 & 0.036\\
MgWide     & 0.285 & 0.002 & 0.400 & 0.025\\
FeI+II+CrI & 0.351 & 0.010 & 0.275 & 0.108\\
2609/2660  & 0.883 & 0.015 & 0.384 & 0.165\\
MgII2800   & 0.795 & 0.014 & 0.625 & 0.157\\
MgI2852    & 0.349 & 0.008 & 0.332 & 0.087\\
2828/2921  & 0.510 & 0.005 & 0.690 & 0.058\\
FeI3000    & 0.199 & 0.003 & 0.117 & 0.028\\
\hline
\end{tabular}
\smallskip

Adopted redshift (1$+$z) = 1.0093 (NGC~5018);
0.9995 (M32).
\label{tab:fos_ind}
\end{table}

While the overall UV spectral energy distributions (SEDs) of M32 and
NGC~5018 are very similar over a wide range of size scales, it is also
clear from Table~1 and from the analysis of Leonardi \& Worthey (2000)
that the stellar population of NGC~5018 is likely of somewhat
different metallicity than that of M32. However, as 3 Gyr-old stellar
populations are dominated by the F-stars that define their main
sequences, metallicity is a side issue in the match of the SEDs; it is
the 3 Gyr-old ages of these two galaxies that dominate their SEDs.

\section{Conclusions}

This work shows that the stellar populations of the gE galaxy
NGC~5018 and the dE galaxy M32 are of similar age ($\sim$3 Gyr-old),
and this relatively young stellar population is spread rather
uniformly within the central few kpc of NGC~5018.  This evidence,
coupled with the rather detailed evidence that indicates that NGC~5018
underwent a recent, major merger, strongly suggests that the young
stellar population we see in this galaxy was the result of that major
merger. As such, it is likely that NGC~5018 is the older brother of
NGC 7252, which Schweizer (e.g, Schweizer 1982) has shown is at the
tail-end of a major merger that is transforming its components into
a gE galaxy now.  As such, our observations raise a number of
questions that bear on how such a major merger forms a gE galaxy a few
billion years after the event. Chief among these questions is the
issue of any evidence of an older stellar population in NGC~5018, or
were literally most of the stars we see formed in the merger event?
NGC~5018 is a galaxy whose stellar populations at all radii are worth
a detailed study to gain a better understand how gE galaxy are formed
via mergers today.

\begin{acknowledgements}
MC acknowledges support from a VENI grant awarded by the Netherlands Organization of Scientific Research
(NWO).
\end{acknowledgements}

\end{document}